\documentclass[twocolumn,pra,superscriptaddress]{revtex4}
\usepackage{amsfonts}
\usepackage{amssymb}
\usepackage{amsmath}
\usepackage{epsfig}
\usepackage{color}
\usepackage{graphics, graphicx}
\usepackage{bbold}
\usepackage{psfrag}
\usepackage{mathcomp}
\usepackage{subfigure}
\usepackage{verbatim}
\usepackage[colorlinks,citecolor=blue]{hyperref}

\makeatletter

\newcommand{\Rmnum}[1]{\expandafter\@slowromancap\romannumeral #1@}
\makeatother

\begin{document}

\title{Topological pumping of a single magnon in a one-dimensional
       spin-dependent optical superlattice}
\author{Feng Mei}
\email{meifeng@sxu.edu.cn}
\affiliation{State Key Laboratory of Quantum Optics and Quantum Optics Devices, Institute
of Laser Spectroscopy, Shanxi University, Taiyuan, Shanxi 030006, China}
\affiliation{Collaborative Innovation Center of Extreme Optics, Shanxi
University,Taiyuan, Shanxi 030006, China}
\author{Gang Chen}
\email{chengang971@163.com}
\affiliation{State Key Laboratory of Quantum Optics and Quantum Optics Devices, Institute
of Laser Spectroscopy, Shanxi University, Taiyuan, Shanxi 030006, China}
\affiliation{Collaborative Innovation Center of Extreme Optics, Shanxi
University,Taiyuan, Shanxi 030006, China}
\author{Liantuan Xiao}
\affiliation{State Key Laboratory of Quantum Optics and Quantum Optics Devices, Institute
of Laser Spectroscopy, Shanxi University, Taiyuan, Shanxi 030006, China}
\affiliation{Collaborative Innovation Center of Extreme Optics, Shanxi
University,Taiyuan, Shanxi 030006, China}
\author{Suotang Jia}
\affiliation{State Key Laboratory of Quantum Optics and Quantum Optics Devices, Institute
of Laser Spectroscopy, Shanxi University, Taiyuan, Shanxi 030006, China}
\affiliation{Collaborative Innovation Center of Extreme Optics, Shanxi
University,Taiyuan, Shanxi 030006, China}
\date{\today }

\begin{abstract}
Topological pumping of ultracold atomic gases has recently been
demonstrated in two experiments (Nat. Phys. 12, 296; 12, 350 (2016)).
Here we study the topological pumping of a
single magnon in a dynamically controlled spin-dependent optical
superlattice. When the interaction between atoms is strong, this
system supports a dynamical version of topological magnon insulator
phase. By initially putting a single magnon in the superlattice and slowly
varying the dynamical controlled parameter over one period, the shift of the
magnon density center is quantized and equal to the topological Chern
number. Moreover, we also find that the direction of this quantized shift is
entanglement-dependent. Our result provides a route for realizing topological pumping of
quasiparticles in strongly correlated ultracold atomic system and for studying the interplay between topological pumping and quantum entanglement.
\end{abstract}

\maketitle

\section{Introduction}

In 1983, Thouless studied the transport of particles in a one-dimensional
slowly varying periodic potential \cite{Thouless}. Such varying makes the
Hamiltonian of the system experience a cyclic evolution. Suppose the
particles are initially prepared in the ground band of the potential. When
the Hamiltonian evolves adiabatically with time and returns to the original
point after a period, the number of transported particles is equal to the
topological Chern number associated with the topology of the ground band.
This pumping depends only on the geometric properties of the pump cycle and
is robust to disorder and interaction effects \cite{Thouless}. Although the
Thouless pumping was predicted more than thirty years ago, it is still a
challenge to be realized in solid state systems.

In the past years, investigating topological and geometric pumping with
tunable ultracold atomic system has attracted a lot of interests \cite{Zhang,Gong,Wang,Mei,Wei,Cooper,Zhai,Fazio,Xu,Qin,Price,WangBotao,Lu,TPFermion,TPBoson}.
Specifically, two experimental groups have recently reported the realization
of topological Thouless pumping with ultracold fermionic and bosonic atoms
trapped in spin-independent optical superlattices \cite{TPFermion,TPBoson}. Populating the
topologically nontrivial ground band is archived by placing the Fermi energy
in the band gap \cite{TPFermion} or by preparing the bosonic atoms into Mott
insulator in the ground band \cite{TPBoson}. The shift of the center-of-mass
of the atomic cloud during a pump cycle amounts to the Chern number.
Moreover, two-dimensional topological pumping protected by the second Chern
number has also been demonstrated experimentally in a two-dimensional
optical superlattice \cite{2DTP}. However, all these studies focus on
topological pumping of an atomic gas and requires preparing this gas in the
ground band.

On the other hand, ultracold atoms trapped in optical lattices can also be
employed to realize spin chain models and explore magnonic states \cite%
{DuanSCM,SCM2003b,SCM2004a,SCM2004b,DMa,DMb,DMc}. Via single-site- and time-resolved
technologies in optical lattices, based on mimicked spin chains, recent
ultracold atoms experiments have successfully observed the quantum dynamics
of single- and two-magnon states \cite%
{Fukuhara2013a,Fukuhara2013b,Fukuhara2015}. Different from neutral atoms,
magnons are bosonic quasiparticle excitations around the ground state of
strongly correlated quantum spin chain models \cite%
{Fukuhara2013a,Fukuhara2013b,Fukuhara2015}, which are important for the
development of spintronics devices. Meanwhile, the concept of topology has
been further expanded to magnonic systems \cite{TMI2013,TMI2014}. Searching
topological magnon insulator and semimetal phases in solid state systems has
also attracted much attention \cite%
{Lee2015,Fiete2017,CG2016,Mertig2016,FC2017}.

In this paper, we investigate the topological magnon insulator phase and the
topological magnon pumping in a dynamically controlled spin-dependent
optical superlattice system. The tight-binding form of this dynamical system
is described by the Rice-Mele-Bose-Hubbard model with a dynamical parameter $%
\theta$. When the on-site interaction is strong, the system can represent a
spin chain with two spins per unit cell, where the intercell and intracell
couplings are different and the on-site energies are spin-dependent. In the
process of adiabatically varying $\theta$, we show that a dynamical version
of topological magnon insulator phase characterized by the Chern number can
emerge. After preparing a single-magnon state in the system, we demonstrate
that quantized topological pumping of a single magnon can be implemented by
slowly tuning $\theta$ over one period. We also exhibit that the pumping
direction depends on the internal spin entanglement configuration in the
initial single-magnon state. Compared with previous Thouless pumping
experiments \cite{TPFermion,TPBoson}, our work studies the topological
pumping of a single quasiparticle in a strongly correlated system, where the
initial ground band population can be easily prepared. In addition, we find
that quantum entanglement can play an important role in the topological
pumping. Finally, we also show that this topological pumping can be
efficiently detected based on a parallel state preparation and detection
strategy.

This paper is organized as follows. In Sec.~II, we present the dynamically
controlled spin-dependent optical superlattice system. In Sec.~III, we show
a dynamical version of topological magnon insulator phase can be generated.
In Sec.~IV, we investigate the entanglement-dependent topological pumping of
a single magnon in a strongly correlated ultracold atomic system. In Sec. V,
we briefly discuss how to prepare the initial state and realize the parallel
topological pumping. In Sec.~VI, we give a summary for the main results in
this work.

\section{Rice-Mele-Bose-Hubbard model}

We consider ultracold bosonic $^{87}$Rb atoms trapped in a one-dimensional
spin-dependent optical superlattice. Each atom is assumed to have two
relevant internal states labeled respectively by the spin index $|\downarrow
\rangle =|F=1,m_{F}=-1\rangle $ and $|\uparrow \rangle =|F=2,m_{F}=-2\rangle
$. Such spin-dependent optical superlattice has been realized in experiments
by superimposing two standing optical waves \cite{Yuan,Bloch}. The
corresponding potential $V_{\sigma }(x)=V_{l\sigma }\sin
^{2}(k_{1}x)+V_{s\sigma }\sin ^{2}(2k_{1}x+\varphi _{\sigma })$, where the
potential depths $V_{(l,s)\sigma }$ and the laser phase $\varphi _{\sigma }$
can be varied by changing the laser power and the optical path difference.
For sufficiently deep optical lattice potential and low temperatures, this
optical superlattice system can be described by the Rice-Mele-Bose-Hubbard
model
\begin{eqnarray}
H &=&H_{0}+V,  \notag \\
H_{0} &=&-\sum_{x=1}^{N}\sum_{\sigma=\uparrow,\downarrow}(J_{1}\hat{a}_{i,\sigma }^{\dag }\hat{b}%
_{i,\sigma }+J_{2}\hat{b}_{i,\sigma }^{\dag }\hat{a}_{i+1,\sigma }+\text{H.c.%
})  \notag \\
&+&\Delta \sum_{i}(\hat{a}_{i\uparrow }^{\dag }\hat{a}_{i\uparrow }-\hat{b}%
_{i\uparrow }^{\dag }\hat{b}_{i\uparrow }-\hat{a}_{i\downarrow }^{\dag }\hat{%
a}_{i\downarrow }+\hat{b}_{i\downarrow }^{\dag }\hat{b}_{i\downarrow }),
\notag \\
V &=&\sum_{x=1}^{N}\sum_{\sigma=\uparrow,\downarrow}\frac{U}{2}(\hat{a}_{i,\sigma }^{\dag }\hat{a}%
_{i,\sigma }^{\dag }\hat{a}_{i,\sigma }\hat{a}_{i,\sigma }+\hat{b}_{i,\sigma
}^{\dag }\hat{b}_{i,\sigma }^{\dag }\hat{b}_{i,\sigma }\hat{b}_{i,\sigma })
\notag \\
&+&U\sum_{i}(\hat{a}_{i\uparrow }^{\dag }\hat{a}_{i\uparrow }\hat{a}%
_{i\downarrow }^{\dag }\hat{a}_{i\downarrow }+\hat{b}_{i\uparrow }^{\dag }%
\hat{b}_{i\uparrow }\hat{b}_{i\downarrow }^{\dag }\hat{b}_{i\downarrow }),
\end{eqnarray}%
where $\hat{a}_{i,\sigma }^{\dag }$ ($\hat{b}_{i,\sigma }^{\dag }$) is the
spin-dependent creation operator associated with the lattice site $a_{i}$ ($%
b_{i}$) in the $i$-th unit cell, $J_{1,2}=J\mp \delta J$ are the alternating
tunneling amplitudes, $\Delta $ is the spin-dependent staggered on-site
energy, $U$ is the on-site interaction and $N$ is the unit cell number.
The interaction between atoms in the same or different spin state is assumed to be same. In
the absence of interaction, the above model corresponds to the spinful
Rice-Mele model \cite{RMModel}. The spinless Rice-Mele model and its
topological features have been experimentally studied in a one-dimensional
optical superlattice \cite{CASSH}.

In the strong interaction case, $V$ (the hopping and on-site energy terms)
can be seen as a perturbation to $H_{0}$ (the on-site interaction term). We
label the ground and excitation state subspaces spanned by the eigenstates
of $H_{0}$ as $\mathcal{P}$ and $\mathcal{Q}$. For unit filling, the ground
and excitation state subspaces associated with one unit cell can be written
as
\begin{eqnarray}
\mathcal{P} &=&\{|\uparrow, \uparrow \rangle ,|\uparrow, \downarrow \rangle
,|\downarrow, \uparrow \rangle ,|\downarrow, \downarrow \rangle \},  \notag \\
\mathcal{Q} &=&\{|\uparrow \uparrow ,0\rangle ,|0,\uparrow \uparrow \rangle
,|\uparrow \downarrow ,0\rangle ,  \notag \\
&&|0,\uparrow \downarrow \rangle ,|\downarrow \downarrow ,0\rangle
,|0,\downarrow \downarrow \rangle \},
\end{eqnarray}%
with the eigenenergies as $E_{g}=0$ and $E_{e}=U$. The corresponding
projective operators for the above two subspaces are defined as
\begin{eqnarray}
\hat{P} &=&\sum_{|j\rangle \in \mathcal{P}}|j\rangle \langle j|,  \notag \\
\hat{Q} &=&\sum_{|k\rangle \in \mathcal{Q}}|k\rangle \langle k|.
\end{eqnarray}%
Based on the Schrieffer-Wolf transformation \cite{SWT}, the low energy
effective Hamiltonian up to second order can be formulated as
\begin{equation}
H_{e}=\hat{P}H_{0}\hat{P}+\hat{P}V\hat{P}+\frac{\hat{P}V\hat{Q}V\hat{P}}{%
E_{g}-E_{e}}.  \label{Heff}
\end{equation}%
It is easy to check that the first term $\hat{P}H_{0}\hat{P}=0$ in our
model. For the intra-cell coupling in the last two terms in Eq.~(\ref{Heff}%
), after a straightforward calculation, we find that
\begin{equation}
\hat{P}V\hat{P}=\Delta (|\uparrow, \downarrow \rangle \langle \uparrow,
\downarrow |-|\downarrow, \uparrow \rangle \langle \downarrow, \uparrow |),
\label{effB}
\end{equation}%
and
\begin{eqnarray}
\frac{\hat{P}V\hat{Q}V\hat{P}}{E_{g}-E_{e}}=-&&\frac{2J_{1}^{2}}{U}%
(|\uparrow, \downarrow \rangle \langle \uparrow, \downarrow |+|\uparrow,
\downarrow \rangle \langle \downarrow, \uparrow |+\text{H.c.})  \notag \\
&-&\frac{4J_{1}^{2}}{U}(|\uparrow, \uparrow \rangle \langle \uparrow, \uparrow
|+|\downarrow, \downarrow \rangle \langle \downarrow, \downarrow |),
\end{eqnarray}%
where H.c. is the Hermitian conjugate. Similarly, the effective Hamiltonian
with respect to the inter-cell coupling can be derived. Now we introduce the
pauli operators $\hat{S}^{x}=|\uparrow \rangle \langle \downarrow
|+|\downarrow \rangle \langle \uparrow |$, $\hat{S}^{y}=-i|\uparrow \rangle
\langle \downarrow |+i|\downarrow \rangle \langle \uparrow |$, and $\hat{S}%
^{z}=(\hat{n}_{\uparrow }-\hat{n}_{\downarrow })/2$. In terms of these Pauli
operators, the total effective Hamiltonian can be derived as
\begin{eqnarray}
H_{e} &=&\sum_{i}[J_{e1}(\hat{S}_{a_{i}}^{x}\hat{S}_{b_{i}}^{x}+\hat{S}%
_{a_{i}}^{y}\hat{S}_{b_{i}}^{y})+J_{z1}\hat{S}_{a_{i}}^{z}\hat{S}_{b_{i}}^{z}
\notag \\
&+&J_{e2}(\hat{S}_{b_{i}}^{x}\hat{S}_{a_{i+1}}^{x}+\hat{S}_{b_{i}}^{y}\hat{S}%
_{a_{i+1}}^{y})+J_{z2}\hat{S}_{b_{i}}^{z}\hat{S}_{a_{i+1}}^{z}]  \notag \\
&+&\sum_{i}\Delta (\hat{S}_{a_{i}}^{z}-\hat{S}_{b_{i}}^{z}),  \label{DSPC}
\end{eqnarray}%
where $J_{e1,e2}=-J_{1,2}^{2}/U$ and $J_{z1,z2}=-4J_{1,2}^{2}/U$ are the
effective spin superexchange couplings. Such alternating couplings can be
rewritten as $J_{e1,e2}=(J\mp \delta J)^{2}/U=J_{e}\mp \delta J_{e}$.

\begin{figure}[t]
\includegraphics[width=8cm,height=6cm]{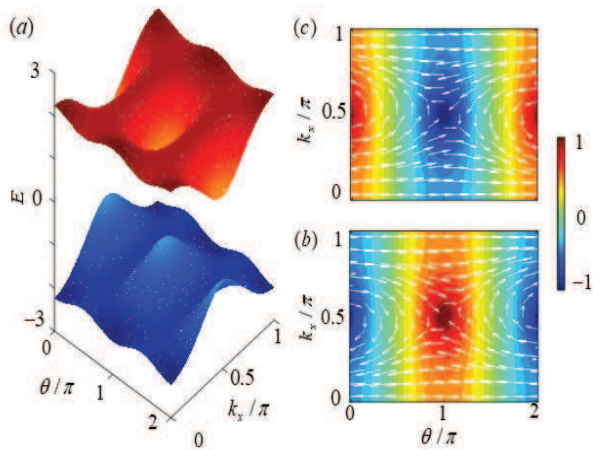}
\caption{(a) The single-magnon energy spectrum with two magnon bands. The
configuration of the unit vector field $\mathbf{n}$ in the first Brillouin
zone for the lower (b) and upper (c) magnon bands. The parameters are chosen
as $J_e=J_p=J$. $J$ is used as energy unit in this work.}
\label{Fig1}
\end{figure}

\section{Dynamical version of topological magnon insulator}

In the present work, only single spin-up excitation is considered in the
above optical superlattice system, which is also called as single-magnon
excitation in condensed matter physics \cite{Fukuhara2013a}. We will study
the topological feature of Eq.~(\ref{DSPC}) in the single-magnon subspace.
Based on the Matsubara-Matsuda mapping \cite{MM}, the above effective spin
model in the single-magnon space can be rewritten into the following magnon
Hamiltonian
\begin{eqnarray}
H_{m} &=&\sum_{i}(J_{e1}\hat{m}_{a_{i}}^{\dag }\hat{m}_{b_{i}}+J_{e2}\hat{m}%
_{b_{i}}^{\dag }\hat{m}_{a_{i+1}}+\text{H.c.})  \notag \\
&+&\sum_{i}\Delta (\hat{m}_{a_{i}}^{\dag }\hat{m}_{a_{i}}-\hat{m}%
_{b_{i}}^{\dag }\hat{m}_{b_{i}}),  \label{Hm}
\end{eqnarray}%
where $\hat{m}_{a_{i}(b_{i})}^{\dag }=|\uparrow \rangle
_{a_{i}(b_{i})}\langle \downarrow |$ is the magnon creation operator
associated with the lattice site $a_{i}(b_{i})$ in the $i$-th unit cell. $%
|G\rangle =|\downarrow \downarrow \cdots \downarrow \downarrow \rangle $ can
be seen as the magnon vacuum state. Experimentally, one can adiabatically
tune the optical lattice potential to vary $(\delta J,\Delta )$ and modulate
$(\delta J_{e},\Delta )$ in a close circle \cite%
{TPFermion,TPBoson,2DTP,Bloch}. In this way, $\big(\delta J_{e},\Delta )$
can be parameterize as $(J_{p}\sin (\theta ),J_{p}\cos (\theta )\big)$,
where $\theta $ is a dynamical parameter. Then the spin superexchange
couplings and on-site energy offset can be written as $J_{e1,e2}=J_{e}\mp
J_{p}\sin (\theta )$ and $\Delta =J_{p}\cos (\theta )$. To investigate the
topological features of the magnon Hamiltonian (\ref{Hm}), we write it in
the momentum space as $H=\sum_{k_{x}}\hat{m}_{k_{x}}^{\dag }\hat{h}%
(k_{x},\theta )\hat{m}_{k_{x}}$, where $\hat{m}_{k_{x}}=(\hat{m}_{a_{k_{x}}},%
\hat{m}_{b_{k_{x}}})^{T}$. Specifically, the momentum density is written as
\begin{equation}
\hat{h}(k_{x},\theta )=h_{x}\hat{\sigma}_{x}+h_{y}\hat{\sigma}_{y}+h_{z}\hat{%
\sigma}_{z},
\end{equation}%
where $h_{x}=2J_{e}\cos (k_{x})$, $h_{y}=2J_{p}\sin (\theta )\sin (k_{x})$
and $h_{z}=J_{p}\cos (\theta )$. $\hat{\sigma}_{x,y,z}$ are the Pauli
matrixes spanned by $\hat{m}_{a_{k_{x}}}$ and $\hat{m}_{b_{k_{x}}}$.

Interestingly, we can construct a two-dimensional artificial Brillouin zone
based on the momentum $k_{x}\in (0,\pi ]$ and the dynamical parameter $%
\theta \in (0,2\pi ]$. The single-magnon energy spectrum is plotted in Fig. %
\ref{Fig1}(a), which has two magnon bands. The topological features of these
two magnon bands are characterized by the Chern numbers. Based on a mapping
from the momentum space to an unit sphere, i.e., $T^{2}\rightarrow S^{2}$,
the Chern number can be defined as \cite{TOPOKane,TOPOZhang}
\begin{equation}
C=\frac{1}{4\pi }\int \int dk_{x}d\theta (\partial _{k_{x}}\mathbf{n}\times
\partial _{\theta }\mathbf{n})\cdot \mathbf{n},
\end{equation}%
where the unit vector field $\mathbf{n}=(h_{x},h_{y},h_{z})/h$ with $h=\sqrt{%
h_{x}^{2}+h_{y}^{2}+h_{z}^{2}}$. The integrand $\mathbf{n}\times \partial
_{\theta }\mathbf{n}\cdot \mathbf{n}$ is simply the Jacobian of this
mapping. Its integration is a topological winding number giving the total
area of the image of the Brillioun zone $T^{2}$ on $S^{2}$ \cite%
{TOPOKane,TOPOZhang}. It means that, when $(k_{x},\theta )$ wraps around the
entire first Brillouin zone $T^{2}$, this winding number is equal to the
number of times the vector $\mathbf{n}$ wraps around the unit sphere $S^{2}$%
, which is independent of the details of the band structure parameters.

In Figs.~\ref{Fig1}(b) and \ref{Fig1}(c), we plot both the unit vector
configurations $(n_{x},n_{y})$ and the contours of $n_{z}$ for the lower and
upper magnon bands, respectively. The results show that, for the lower
(upper) magnon band, the unit vector $\mathbf{n}$ starts from the noth
(south) pole at the Brillouin zone center and ends at the south (north) pole
at the Brillouin zone boundary after wrapping around the unit sphere once.
Thus the Chern number corresponding to the lower (upper) magnon band is
derived as $C_{l}=1$ ($C_{u}=-1$). Since $\theta $ is a periodic dynamical
parameter introduced to construct the first Brillouin zone, the above
nontrivial Chern number values yield a dynamical version of the topological
magnon insulator phase.

\section{Entanglement-dependent topological magnon pumping}

In the following, based on the above dynamically controlled topological
magnon bands, we will demonstrate that entanglement-dependent topological
pumping of a single magnon can be implemented. This is done by adiabatically
tuning the parameter $\theta=\Omega t+\theta_0$ over one period, where $%
\Omega$ is the modulation frequency and $\theta_0$ is the initial phase.

\begin{figure}[t]
\includegraphics[width=7cm,height=6.5cm]{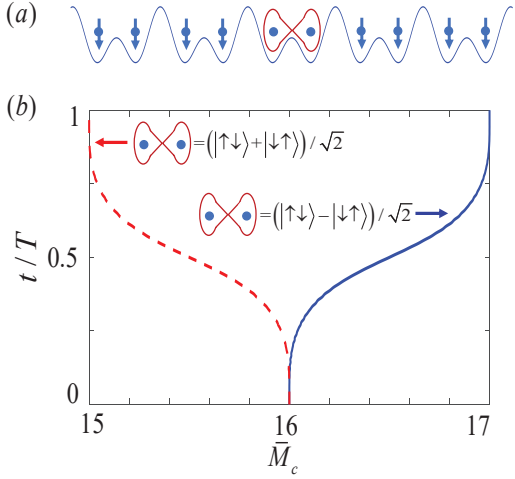}
\caption{(a) The schematic diagram of a one-dimensional optical superlattice
system prepared in a single-magnon state. (b) The change of the magnon
density center $\bar{M}_c$ versus time for different initial spin entangled
state. The parameters are chosen as $J_e=J_p=J$ and $\Omega=0.1J$. The unit
cell number in the optical superlattice is $N=62$.}
\label{Fig2}
\end{figure}

At the initial time, we assume the whole optical superlattice system
consists of series of independent double wells by tuning the optical
superlattice potential to make $J_{e2}=0$. This is equivalent to require the
initial periodic parameter $\theta (t=0)=\theta _{0}=-\text{arcsin}%
(J_{e}/J_{p})$. Suppose the initial system stays in the magnon vacuum state $%
|G\rangle =|\downarrow \downarrow \cdots \downarrow \downarrow \rangle $.
Note that the Hamiltonian for the single-magnon excitation in each double
well has two eigenstates
\begin{equation}
|\chi _{l,u}\rangle =\frac{1}{\sqrt{2}}\left( |\uparrow \downarrow \rangle
\mp |\downarrow \uparrow \rangle \right) ,
\end{equation}%
which are two different Bell states. As shown in Fig. \ref{Fig2}(a), suppose
that one of middle double wells is prepared into the Bell states $|\chi
_{l}\rangle $, while the other qubits stay in the ground state. The initial
state of the system then can be described by $|\psi _{l}\rangle =|\downarrow
\downarrow \cdots \chi _{l}\cdots \downarrow \downarrow \rangle $. Actually,
one can find that $|\psi _{l}\rangle $ is just the Wannier function
corresponding to the lower magnon bands. The reason is that the coupling
between nearest neighbor double wells is zero and so the Wannier function
only localizes the middle double well.

After having prepared the initial single-magnon state, the parameter $\theta$
is adiabatically modulated from $\theta(t=0)$ to $\theta(t=T=2\pi/\Omega)$,
the single-magnon wave packet will experience an adiabatic transfer. We
employ a magnon density center to monitor this transfer. Specifically, the
operator for such center is defined as
\begin{equation}
\hat{M_c}=\sum^{N}_{x=1} x(\hat{P}_{a_x}+\hat{P}_{b_x}),
\end{equation}
where $\hat{P}_{a_x(b_x)}=\hat{m}_{a_x(b_x)}^\dag\hat{m}_{a_x(b_x)}=|%
\uparrow\rangle_{a_x(b_x)}\langle \uparrow|$ is the magnon density in the
lattice site $a_x$ ($b_x$). Then, when the parameter $\theta $ is
adiabatically modulated, the corresponding magnon density center can be
written as
\begin{eqnarray}
\bar{M}_{c}(\theta)&=&\langle \psi _{l}(\theta)|\hat{M_c}|\psi
_{l}(\theta)\rangle  \notag \\
&=&\frac{1}{2\pi }\int dk_{x}i\langle u_{k_{x},\theta,l}|\partial
_{k_{x}}|u_{k_{x},\theta,l}\rangle  \notag \\
&=&\frac{1}{2\pi }\int dk_{x}A_{l}(k_{x},\theta ).
\end{eqnarray}
It turns out that the magnon density center is linked to the Berry
connection $A_{l}(k_{x},\theta )=i\langle u_{k_{x},\theta ,l}|\partial
_{k_{x}}|u_{k_{x},\theta ,l}\rangle $. Therefore, the magnon density center
depends on the gauge choice of the Bloch state. However, the change of the
magnon density center is gauge invariant and thus can be well defined.

Suppose the periodic parameter $\theta $ is changed continuously from $\theta
_{i}$ to $\theta _{f}$. The resulting magnon density center shift is
\begin{equation}
\bar{M}_{c}\left( \theta _{f}\right) -\bar{M}_{c}\left( \theta _{i}\right) =%
\frac{1}{2\pi }\int dk_{x}\big(A_{l}(k_{x},\theta _{f})-A_{l}(k_{x},\theta
_{i})\big).  \label{MS}
\end{equation}%
By means of the Stokes theorem, the formula (\ref{MS}) can be rewritten as
an integral of the Berry curvature $F_{l}(k_{x},\theta )$ over the surface
spanned by $k_{x}$ and $\theta $, where $F_{l}(k_{x},\theta )=\nabla \times
A_{l}(k_{x},\theta )=i\left( |\langle \partial _{\theta }u_{k_{x},\theta
,l}|\partial _{k_{x}}u_{k_{x},\theta ,l}\rangle -\text{c.c.}\right) $. For a
periodic cycle, $\theta _{f}=\theta _{i}+2\pi $, $H(\theta _{i})=H(\theta
_{f})$, and the change of the magnon center over one cycle is given by the
integral of the Berry curvature over the torus $\{k_{x}\in (0,\pi ],\theta
\in (0,2\pi ]\}$. It is easy to check that the magnon center shift in this
case is just the Chern number of the lower magnon band, i.e.,
\begin{eqnarray}
\bar{M}_{c}(\theta _{f})-\bar{M}_{c}(\theta _{i}) &=&\frac{1}{2\pi }%
\int_{k_{x}}\int_{\theta }dk_{x}d\theta \nabla \times A_{l}(k_{x},\theta )
\notag \\
&=&\frac{1}{2\pi }\int_{k_{x}}\int_{\theta }dk_{x}d\theta
\,F_{l}(k_{x},\theta )  \notag \\
&=&C_{l}.
\end{eqnarray}%
Similarly, if the initial single-magnon excitation in the middle double well
is prepared in the Bell state $|\chi _{u}\rangle $, after tuning the
parameter $\theta $ over one period, the shift of the magnon density center
becomes $C_{u}$. Therefore, the entanglement-dependent topological pumping
of a single magnon is achieved.

The detailed performance of the above topological pumping is numerically
calculated in Fig.~\ref{Fig2}(b). Suppose the two atoms in the middle double
well are prepared into the Bell state $|\chi _{l}\rangle $ ($|\chi
_{u}\rangle $), the numerical results show that the magnon density center is
shifted to the right (left) by one unit cells after one periodic pumping,
which equals the Chern number of the lower (upper) magnon band $C_{l}=1$ ($%
C_{u}=-1$). Interestingly, one can find that different Bell states inside
the initial magnon wave packet gives rise to different quantized topological
pumping. Such process generates an entanglement-dependent topological magnon
pumping, which is different from the Thouless pumping and shows that the
internal entanglement configuration in the transport particle can also
affect the external pumping.

\section{Initial state preparation and parallel topological pumping}

\begin{figure}[t]
\includegraphics[width=7cm,height=9cm]{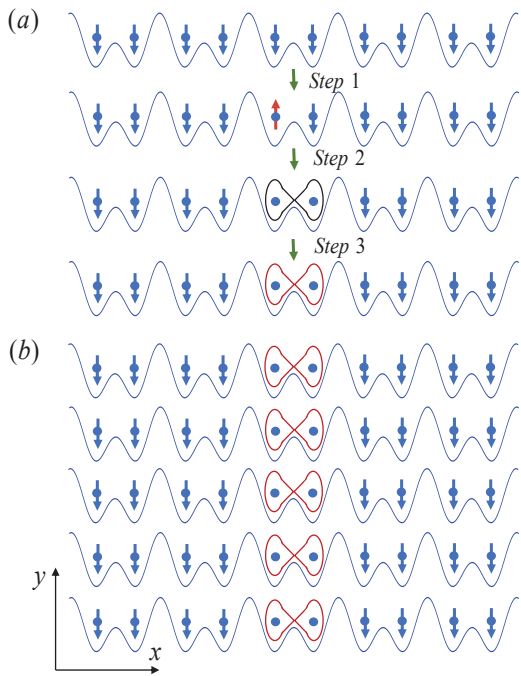}
\caption{(a) The procedure for preparing a spin entangled state in the
middle double well. (b) An array of one-dimensional spin-dependent optical
superlattices for highly efficient parallel entangled state generation and
magnon density detection.}
\label{Fig3}
\end{figure}

The initial single-magnon state with internal spin entanglement can be
prepared based on spin superexchange Hamiltonian and an effective magnetic
field. This method has recently been experimentally demonstrated in a
spin-dependent optical superlattice system \cite{Yuan}. The detailed
preparation procedure is shown in Fig. \ref{Fig3}(a). Initially, the
superlattice potential is tuned to make $J_{e2}=0$ and $\Delta=0$. The
resulted lattice is formed by an array of independent double wells. Suppose
the system is initially prepared in the Mott-insulator regime where each
lattice site has a single atom prepared in the state $|\downarrow \rangle$.
The initial single-magnon state can be prepared via three steps. Step 1:
based on single-site microwave pulse addressing, one of the spins in the
middle double well is flipped into $|\uparrow\rangle$, then the state of the
two spins in the middle double well becomes $|\uparrow\downarrow\rangle$.
Step 2: through a dynamical evolution governed by the spin superexchange
Hamiltonian in Eq.~(\ref{DSPC}) with $\Delta=0$ and evolution time $%
t=\pi/4J_{e1}$, an entangled state $(|\uparrow \downarrow \rangle
-i|\downarrow \uparrow \rangle)/\sqrt{2}$ is generated between the two spins
in the middle double well. Step 3: tuning the spin-dependent optical
superlattice potential to switch on $\Delta$, an effective magnetic field
described by the Hamiltonian shown in Eq.~(\ref{effB}) is created. Via such
magnetic field to modulate the phase of the entangled state, the Bell state $%
(|\uparrow \downarrow \rangle \pm|\downarrow \uparrow \rangle)/\sqrt{2}$ can
be generated. In the whole process, the state of the spins in the other
double wells dose not change. Then the single-magnon state $|\psi
_{l,u}\rangle =|\downarrow\downarrow \cdots \chi _{l,u}\cdots \downarrow
\downarrow \rangle$ is prepared.

In experiment, the topological pumping proposed in this work can be
efficiently measured based on a strategy using parallel state preparation
and detection. As shown in Fig. \ref{Fig3}(b), a two-dimensional degenerate
Bose gas of $^{87}$Rb atoms is prepared in a two-dimensional optical lattice
potential $V(x,y)=V(x)+V(y)$, where $V(x)=V_{\uparrow }(x)+V_{\downarrow }(x)
$ is the state-dependent superlattice potential in the $x$ direction and $%
V(y)=V_{y}\sin ^{2}(2k_{1}y)$ is a spin-independent potential in the $y$
direction. When $V_{y}$ is tuned to be vary large, the hopping along the $y$
direction can be ignored. Then $V(x,y)$ creates an array of independent
one-dimensional optical superlattices along the $x$ direction. In the
Mott-insulator regime, this system can be seen as an array of parallel
one-dimensional spin chains described by Eq.~(\ref{DSPC}). Via an addressing
beam profile in form of a line reported in the recent single-magnon
experiment \cite{Fukuhara2013a}, the left spin down state in each middle
double well of the optical superlattice along the $x$ direction can be
simultaneously flipped into the spin up state, then each superlattice system
has been put into a single magnon. After that, based on the above
entanglement generation procedure and its parallel version, each spin chain can be prepared into the single-magnon state $%
|\psi _{l,u}\rangle $. Finally, through tuning $\theta $ over one period in
each spin chain, we can realize a parallel topological pumping of a single
magnon in this two-dimensional optical lattice system. In this case, the
magnon density can be efficiently measured by averaging the data extracted
from the magnon density measurements in all spin chains \cite%
{Fukuhara2013a,Fukuhara2013b}.

\section{Summary}

In summary, we have realized a dynamical version of the topological magnon
insulator phase and also topological pumping of a single magnon in a
spin-dependent optical superlattice system. Different from previous
experiments implementing topological Thouless pumping of an ultracold atomic
gas, our work focused on topological pumping of a single quasiparticle in a
strongly correlated spinful Bose-Hubbard system. Furthermore, we have also
found that the shift of pumping direction is entanglement-dependent.
Our model is also compatible with the recent optical lattice experiments on magnons \cite%
{Fukuhara2013a,Fukuhara2013b,Fukuhara2015}. Our result opens a prospect for studying topological magnon insulator phase in optical lattice system and also for investigating the interplay between
topological pumping and quantum entanglement and interaction effect.

\section{Acknowledgements}

This work is supported by the National Key R\&D Program of China
(2017YFA0304203); Natural National Science Foundation of China (NSFC)
(11604392, 11674200, 11434007); Changjiang Scholars and Innovative Research
Team in University of Ministry of Education of China (PCSIRT)(IRT\_17R70);
Fund for Shanxi 1331 Project Key Subjects Construction; 111 Project (D18001).

\end{document}